\documentclass[aps,prl,preprintnumbers,amsmath,amssymb,latexsym,array,enumerate,letter,twocolumn,superscriptaddress]{revtex4-1}

\usepackage{graphicx}
\usepackage{dcolumn}
\usepackage{bm}
\usepackage{epsfig}
\usepackage{gensymb}
\usepackage[dvipsnames]{xcolor}
\usepackage{mathrsfs}  
\usepackage{amsmath}
\usepackage{amssymb}
\usepackage{graphicx,bm}
\usepackage{slashed}
\usepackage[makeroom]{cancel}
\usepackage{color}
 \usepackage{titlesec}
   \titleformat{\section}[runin]
   {\normalfont\bfseries}{\quad\thesection.}{0.5em}{} [.]
  \titlespacing*{\section}{0pt}{0.2\baselineskip}{\baselineskip}

\def\fun#1#2{\lower3.6pt\vbox{\baselineskip0pt\lineskip.9pt
  \ialign{$\mathsurround=0pt#1\hfil##\hfil$\crcr#2\crcr\sim\crcr}}}

\newcommand{\Tr}{{\rm Tr\hskip 2pt}}
\def\lsim{\mathrel{\rlap{\raise 2.5pt \hbox{$<$}}\lower 2.5pt\hbox{$\sim$}}}
\def\gsim{\mathrel{\rlap{\raise 2.5pt \hbox{$>$}}\lower 2.5pt\hbox{$\sim$}}}

\renewcommand{\Re}{{\rm Re\thinspace}}
\renewcommand{\Im}{{\rm Im\thinspace}}

\input epsf

\newcommand{\be}{\begin{equation}}
\newcommand{\ee}{\end{equation}}
\newcommand{\bea}{\begin{eqnarray}}
\newcommand{\eea}{\end{eqnarray}}

\usepackage{color}

\newcommand{\mathscrL}[0]{\mathscr{L}}
\newcommand{\mathscrl}[1]{\mathscrL_{\textrm{#1}}}

\newcommand{\comment}[1]{}

\begin{document}

\preprint{ACFI-T16-27}

\title{ Lepton-Flavored Electroweak Baryogenesis}
\author{Huai-Ke Guo}
\affiliation{ Amherst Center for Fundamental Interactions, Department of Physics, University of Massachusetts Amherst, Amherst, MA 01003, USA }
\affiliation{
CAS Key Laboratory of Theoretical Physics, Institute of Theoretical Physics, \\
Chinese Academy of Sciences, Beijing 100190, China}
\author{Ying-Ying Li}
\affiliation{ Department of Physics, The Hong Kong University of Science and Technology, Clear Water Bay, Kowloon, Hong Kong S.A.R., P.R.C.}
\author{Tao Liu}
\affiliation{ Department of Physics, The Hong Kong University of Science and Technology, Clear Water Bay, Kowloon, Hong Kong S.A.R., P.R.C.}
\author{Michael Ramsey-Musolf}
\affiliation{ Amherst Center for Fundamental Interactions, Department of Physics, University of Massachusetts Amherst, Amherst, MA 01003, USA }
\affiliation{
 Kellogg Radiation Laboratory, California Institute of Technology, Pasadena, CA 91125 USA }
\author{Jing Shu}
\affiliation{
CAS Key Laboratory of Theoretical Physics, Institute of Theoretical Physics, \\
Chinese Academy of Sciences, Beijing 100190, China}
\affiliation{
CAS Center for Excellence in Particle Physics, Beijing 100049, China
}

\begin{abstract}
We explore lepton-flavored electroweak baryogenesis, driven by CP-violation in leptonic Yukawa sector, using the $\tau-\mu$ system in the two Higgs doublet model as an example. This setup generically yields, together with the flavor-changing decay $h\to \tau \mu$, a tree-level Jarlskog-invariant that can drive dynamical generation of baryon asymmetry during a first-order electroweak phase transition and results in CP-violating effect in the decay $h\to \tau\tau$.  We find that the observed baryon asymmetry can be generated in parameter space compatible with current experimental results for the decays $h\to \tau \mu$, $h\to \tau\tau$ and $\tau \rightarrow \mu \gamma$, as well as the present bound on the electric dipole moment of the electron. The baryon asymmetry generated is intrinsically correlated with the CP-violating decay $h\to \tau\tau$ and the flavor-changing decay $h\to \tau\mu$, which thus may serve as ``smoking guns'' to test lepton-flavored electroweak baryogenesis.

\end{abstract}


\maketitle

\noindent{\bfseries Introduction.} Explaining the origin of the  baryon asymmetry of the universe (BAU) is a forefront challenge for fundamental physics. The BAU is characterized by the 
baryon density $n_B$ to entropy $s$ ratio 
\begin{eqnarray}
  Y_B = \frac{n_B}{s} = (8.61 \pm 0.09)\times 10^{-11}\ \ \ \text{\cite{Ade:2015xua}} .
\end{eqnarray}
According to Sakharov~\cite{Sakharov:1967dj}, generation of a non-vanishing $Y_B$ requires three ingredients in the particle physics of the early universe: non-conservation of baryon number (B); C- and CP-violation (CPV); and out of equilibrium dynamics (assuming CPT conservation). While the Standard Model (SM) of particle physics contains the first ingredient in the guise of electroweak sphalerons, it fails with regard to the remaining two. Physics beyond the SM is, thus, essential for successful baryogenesis.

Electroweak baryogenesis (EWBG)\cite{Kuzmin:1985mm} is among the most theoretically well-motivated and experimentally testable scenarios, as it ties BAU generation to electroweak symmetry-breaking (see~\cite{Morrissey:2012db} for a recent review). Extending the SM scalar sector can lead to a first order electroweak phase transition(EWPT), thereby satisfying the out-of-equilibrium condition. Addressing the second Sakharov criterion requires new sources of CPV, as the effect of CPV in the  SM Yukawa sector is suppressed by the small magnitude of the Jarlskog invariant associated with the Cabbibo-Kobayashi-Maskawa (CKM) matrix and by the small quark mass differences relative to the electroweak temperature, $T_\mathrm{EW}\sim 100$ GeV.

It is possible that an extended Yukawa sector may remedy this SM shortcoming. A particularly interesting yet unexplored possibility involves the leptonic Yukawa interactions. 
Phenomenologically, the report by the CMS collaboration of a signal for the charged lepton flavor violating (CLFV) Higgs boson decay $h\to\tau\mu$ (2.4 $\sigma$ significance)\cite{Khachatryan:2015kon} hints at a possible richer leptonic Yukawa sector~\cite{CMS:13Tev},
though the ATLAS collaboration observes no evidence for this  decay mode\cite{Aad:2016blu}. Should an extended leptonic Yukawa sector exist, then the accompanying new CPV phases may provide sources for EWBG that do not suffer from the suppression associated with SM quark Yukawa sector.

Motivated by these considerations, we study the viability of \lq\lq lepton flavored EWBG", a scenario that relies on both CLFV and leptonic CPV. For concreteness, we use a variant of the type III 
two Higgs doublet model (2HDM)~\cite{Branco:2011iw} with generic leptonic Yukawa
textures~\cite{Botella:2014ska} and focus on the $\tau-\mu$ families as an example. For a representative choice of Yukawa texture, we derive the CPV source for the EWBG quantum transport equations~\cite{Riotto:1998zb,Lee:2004we} in terms of the relevant Jarlskog invariant, $\text{Im}J_A$. We then solve these equations, which encode the dynamics of CLFV scattering during the electroweak phase transition, and obtain the BAU as a function of the Yukawa matrix parameters. We also show that the same $\text{Im}J_A$ also generates a CPV coupling of the Higgs boson to $\tau$ leptons at $T=0$, parameterized by a CPV phase $\phi_\tau$. Measurements of CPV asymmetries in $h\to\tau^+\tau^-$, as discussed in Ref.~\cite{Hayreter:2016kyv}, would provide a test of this baryogenesis mechanism. Taking into account present constraints from measurements of $\Gamma(h\to\tau^+\tau^-)$ and limits on $\Gamma(\tau\to\mu\gamma)$ we find that a $\mathcal{O}(10{\degree})$ determination of $\phi_\tau$ would probe this scenario at a significant level.

\noindent{\bfseries Model Setup.} Since our focus is on CPV in the $\mu-\tau$ sector which has no mixing with the first-generation charged leptons, 
we assume the scalar potential to be CP-conserving with parameters chosen to generate a 
strongly first order EWPT~\cite{Dorsch:2013wja}. 
The particle spectrum consists of two CP-even neutral scalars ($h, H$), the neutral CP-odd $A$, and a pair of   
charged scalars $H^{\pm}$, with the lighter $h$ taken to be SM-like. The $SU(2)_L\times U(1)_Y$ invariant weak 
eigenbasis lepton Yukawa interaction is
\begin{eqnarray}
  \mathscrl{Yukawa}^{\text{Lepton}} = - \overline{L^{i}} \left[ Y_{1,ij}\Phi_1 + Y_{2,ij} \Phi_2 \right] e_R^{j} + h.c.,
\label{2hdmYukawa}
\end{eqnarray}
where $\Phi_{1,2}$ are the two Higgs doublets with the same hypercharge, $L^{i}$ and $e_R^{j}$ are left-handed lepton doublet and 
right-handed lepton singlet in weak basis, with the family index $i, j=2, 3$. Then we can uniquely define a Jarlskog invariant 
as the imaginary part of~\cite{Jarlskog:1985ht,Botella:1994cs}:
\begin{eqnarray}
  J_A = \frac{1}{v^2 \mu_{12}^{\text{HB}}} 
  \sum_{a,b,c=1}^2 v_a v_b^{\ast} \mu_{bc} \Tr[ Y_c Y^{ \dagger}_a]\ \ , 
      \label{Jinv}
\end{eqnarray}
with the power of Yukawa coupling (or mass parameter of fermions) product being two. Here $v_a=\sqrt{2} \langle \Phi_a^0\rangle$ is vacuum expectation value (vev) of neutral Higgs fields, $\mu_{ab}$  is the coefficient of $\Phi_a^{\dagger} \Phi_b$ in the potential, and the trace is taken over flavor space. $J_A$ is normalized to be 
a dimensionless quantity by dividing a factor $v^2 \mu_{12}^{\text{HB}}$, where 
\begin{equation}
\mu_{12}^{\text{HB}}= \frac{1}{2}(\mu_{22}-\mu_{11})\sin 2\beta + \mu_{12} \cos 2\beta
\end{equation} 
is a quardratic Higgs coupling defined in ``Higgs basis"~\cite{Botella:1994cs,Branco:2011iw}: $H_1 = \cos\beta \Phi_1 + \sin\beta \Phi_2$; $H_2 = -\sin\beta \Phi_1 + \cos\beta \Phi_2$;  $\langle H_1^0\rangle = v/\sqrt{2} = 174$ GeV;  and $\langle H_2^0\rangle=0$.

The mass matrix for fermions is defined as
\begin{eqnarray}
M = ({v_1 Y_1 + v_2 Y_2})/\sqrt{2} 
\end{eqnarray}
in the weak basis, with a determinant of $M^\dagger M$ or $M$ close to zero (since $m_{\mu} \approx 0$). For illustration, we choose a texture with $Y_{j,22} = Y_{j,23} \equiv 0$,  with $j=1,2$. This immediately yields 
\begin{eqnarray}
\Im(J_A) = - \Im (Y_{1,32} Y_{2,32}^* + Y_{1,33} Y_{2,33}^*) 
\end{eqnarray}
or 
\begin{eqnarray}
\Im(J_A) = - \Im (Y_{1,32} Y_{2,32}^* ) \ ,
\end{eqnarray}
with a further assumption $Y_{1,33}=Y_{2,33}$. The diagonalization condition  $|M_{32}|^2 + |M_{33}|^2 = m_{\tau}^2$
immediately gives $\vert M_{32}\vert \le m_{\tau}$, and 
fixes the value of $|Y_{1,33}| = |Y_{2,33}|$.
Since the proposed mass texture is not invariant under basis transformation of $\Phi_1$ and $\Phi_2$, $\tan\beta= v_2/v_1$ becomes an independent parameter (similar to what happens in type II 2HDM). Thus this setup contains five relevant and independent parameters: $\tan\beta$, $\alpha$ (the mixing angle in the CP-even Higgs sector), $|Y_{2,32}|$, $r_{32} = \vert Y_{1,32} \vert/\vert Y_{2,32} \vert$ and $\Im(J_A)$.

In the mass basis for both fermions and Higgs bosons, the $\tau$ Yukawa interaction is then parameterized as 
\begin{eqnarray}
\label{eq:tautau}
-\frac{1}{v} 
&& \overline{\tau_L} \tau_R  [h (m_{\tau} s_{\beta-\alpha} + N_{\tau\tau} c_{\beta-\alpha} )  \nonumber \\
 &&+H (m_{\tau} c_{\beta-\alpha} - N_{\tau\tau} s_{\beta-\alpha} )  + i A N_{\tau\tau} ] + \mathrm{h.c.},
  \label{}
\end{eqnarray}
where $\beta-\alpha$ is invariant under the basis transformation in Higgs family space~\cite{Gunion:2002zf}. The SM-like Higgs boson $h$ receives two contributions to its coupling. The first one results from its $H_1^0$ component which is aligned with the $\tau$ mass. Another one is related to its $H_2^0$ component which is proportional to $N_{\tau\tau}$, the Yukawa coupling of $H_2^0$ with $\tau$ leptons, with 
\begin{eqnarray}
\text{Re}(N_{\tau\tau}) &=& \frac{v^2 \mu_{12}^{\text{HB}}\text{Re}(J_A)- 
2 \mu_{11}^{\text{HB}} m_{\tau}^2}{2\mu_{12}^{\text{HB}} m_{\tau}} \ , \nonumber \\ 
\text{Im}(N_{\tau\tau}) &=& \frac{v^2 \text{Im}(J_A)}{2m_{\tau}} \ . 
\label{ReImNEtt}
\end{eqnarray}
The CLFV interactions are completely controlled by the Yukawa coupling of $H_2^0$, $N_{\tau\mu}$,
\begin{eqnarray}
  - \frac{N_{\tau\mu}}{v} \overline{\tau_L} \mu_R ( c_{\beta-\alpha} h - s_{\beta-\alpha} H + i A) + \mathrm{h.c.} ,
  \label{eq:clfv}
\end{eqnarray}
With $\tan\beta=1$, the expression in terms of weak basis parameters is given by
\begin{eqnarray}
N_{\tau\mu} = e^{i\delta}\left|N_{\tau\tau}\frac{M_{33}}{M_{32}}\right| .
\label{NEtm}
\end{eqnarray}
Here $\delta$ is an un-physical phase undetermined in the diagonalization procedure which can be removed by field redefinition. 
For later convenience, we also have for $\tan\beta=1$
\begin{eqnarray}
\Re(J_A) = \frac{1}{2} ( |Y_{2,32}|^2 -|Y_{1,32}|^2) + \frac{}{} \frac{2m_\tau^2}{v^2}   \frac{\mu_{11}^{\rm HB}}{\mu_{12}^{\rm HB}} \ .
\end{eqnarray}
Finally the charged Higgs Yukawa interactions  are governed by $-\sqrt{2}/v H^+ \overline{\nu_{L}^i} N_{ij} e_R^j + \mathrm{h.c.}$ .

\begin{figure*}
\centering
  \includegraphics[width=0.7\textwidth]{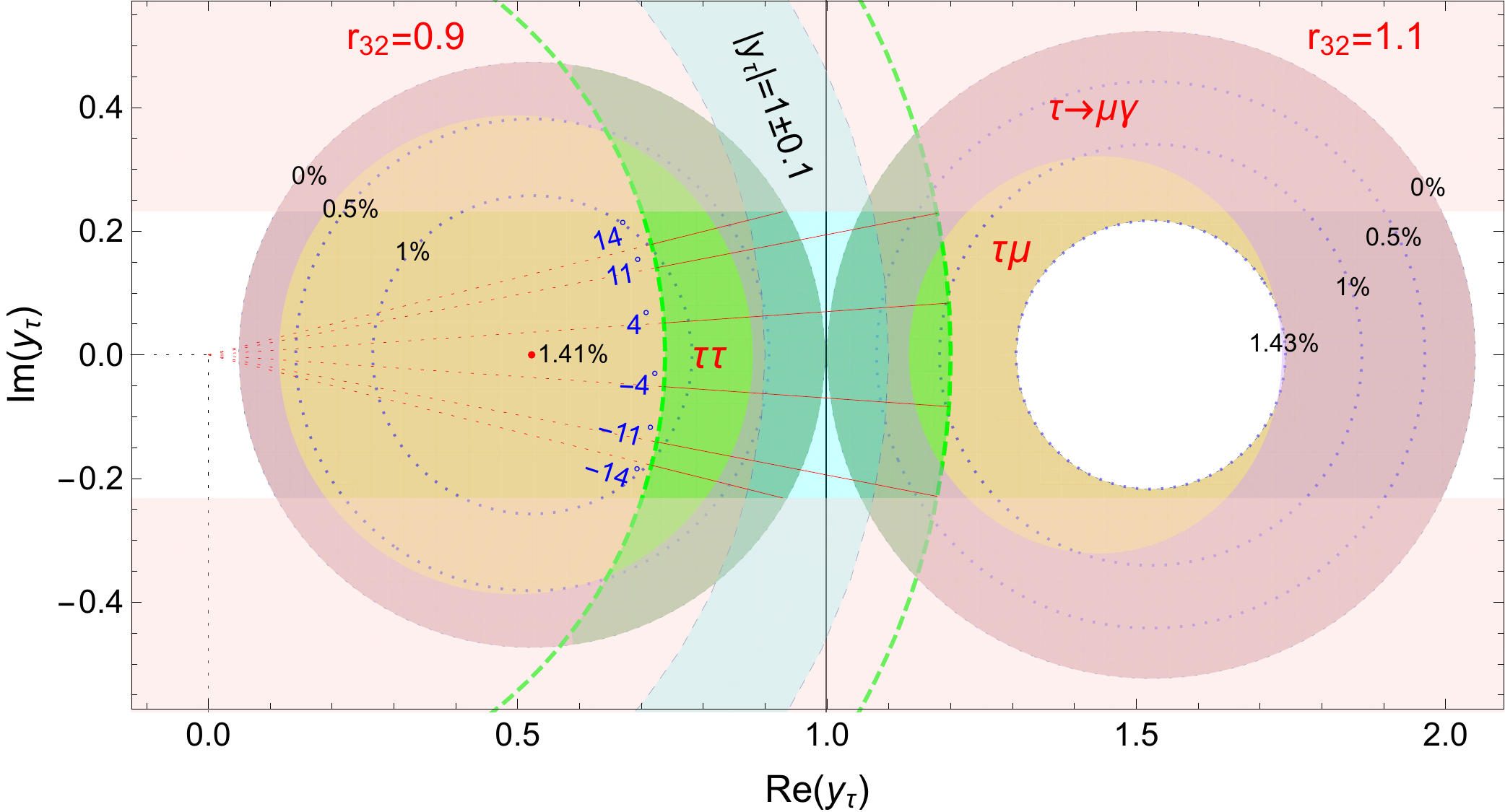}
  \caption{
    \label{CPVTauTau} Theoretical and phenomenological constraints on the Higgs-$\tau$ Yukawa couplings in Eq.~(\ref{htautaukappa}).
    The inner parts of circular regions satisfy the diagonalization constraint $|M_{32}| \le m_{\tau}$ for two representative choices of $r_{32}$, with the outer boundaries giving vanishing $\mathrm{Br}(\tau\rightarrow \mu \gamma)$ and $\Gamma(h\rightarrow \tau \mu)$. The $r_{32} = 0.9$ and $r_{32}=1.1$ regions are separated by the vertical dashed line at 
    $\text{Re}(y_{\tau}) = \sin(0.05+\frac{\pi}{2}) \approx 1$.
Brown regions correspond to non-vanishing $\Gamma(h\rightarrow \tau \mu)$, with different representative values ($1\%$, $0.5\%$ and $0\%$) indicated by circular dashed lines. For $r_{32}=1.1$, the ATLAS 95\% C.L. upper bound of $1.43\%$ is shown, while for $r_{32}=0.9$ a maximum BR of $1.41\%$ can be achieved within the theoretically allowed region. 
Upper limits on $\Gamma(h\rightarrow \tau\tau)$ (95\% C.L.) and  $\mathrm{Br}(\tau\rightarrow \mu\gamma)$(90\% CL) are given by the green and grey regions, respectively.  The region inside the green dashed lines gives the  Higgs signal strength $\mu^{\tau\tau}$ allowed region at 95 \% C.L. without assuming a specific Yukawa texture. The inner light-blue band labelled $|y_\tau| = 1\pm 0.1$ corresponds to the region with a more SM-like $h\bar{\tau}\tau$ coupling. The region giving the observed BAU is indicated by the horizontal pink bands assuming $|\Delta\beta| \le 0.4$) for $\beta-\alpha-\frac{\pi}{2}=0.05$ as discussed in the text. The other relevant 
    parameters are fixed to be $m_H = m_A = m_{H^{\pm}}= 500\text{GeV}$, 
    $v_w=0.05$, $L_W = 2/T$, $D_q=6/T$ and $T=100\text{GeV}$. To guide the eye, the argument of $y_\tau$ is indicated with red-dotted lines. Note, the calculation of baryon asymmetry outside the circular regions could be unreliable due to the breaking of perturbative ``mass insertion''.
}
\end{figure*}

Given the four free parameters left for describing tree-level Yukawa interactions of the $\mu-\tau$ system, we present various phenomenological results (e.g., $h\to \tau\tau,  \tau\mu$ and $\tau\rightarrow \mu\gamma$ constraints) and the BAU analysis 
in terms of the effective $h \bar{\tau}\tau$ coupling~\cite{Berge:2015nua} (see Fig.~\ref{CPVTauTau})
\begin{eqnarray}
  - \frac{m_\tau}{v} [\text{Re}(y_{\tau}) \bar{\tau}\tau + \text{Im}(y_{\tau}) \bar{\tau}i\gamma_5 \tau] h\ \ \,
\label{htautaukappa}
\end{eqnarray}
with benchmark values assigned to $r_{32}$ and $\beta-\alpha$. Here 
\begin{eqnarray}
\Re(y_\tau) &=& s_{\beta -\alpha} + \frac{c_{\beta -\alpha}} {m_\tau} \Re (N_{\tau\tau}) \ , \nonumber \\
\Im(y_\tau) &=& \frac{c_{\beta -\alpha}} {m_\tau} \Im (N_{\tau\tau}) \ .
\end{eqnarray}
Then, the condition $|M_{32}|\le m_{\tau}$ imposes a constraint at the ($\text{Re}(y_{\tau})$, $\text{Im}(y_{\tau})$) plane, allowing only 
a circular region centered at $(\text{Re}(y_{\tau}) = s_{\beta-\alpha}+c_{\beta-\alpha}(1+r^2_{32})/(1-r^2_{32}), 
\text{Im}(y_{\tau}) = 0)$ with a radius $2|c_{\beta-\alpha}r_{32}/(1-r_{32}^2)|$. At its boundary, we have $M_{33}=0$ and hence $N_{\tau\mu} = 0$.  For $r_{32} = 1$, $N_{\tau\tau}$
is purely imaginary, yielding a vertical line at $\text{Re}(y_{\tau}) = s_{\beta-\alpha}$. In Fig.~\ref{CPVTauTau}, we present results in two representative cases: $r_{32} =0.9$ and $r_{32} =1.1$, with  $\beta-\alpha-\frac{\pi}{2}=0.05$.
\medskip

\noindent{\bfseries $\bm{ h\to \tau\tau}$ constraints.}
The decay width for $h \to \tau\tau$ is given by 
\begin{eqnarray}
  \Gamma^{\tau\tau} =\frac{\sqrt{2} G_F m_h{ m_{\tau}^2}}{8\pi} |y_\tau|^2  \ .
  \label{TauTauWidth}
\end{eqnarray}
Experimentally, the ATLAS signal strength is 
$\mu^{\tau\tau}_{\text{ATLAS}}=1.43^{+0.43}_{-0.37}$~\cite{Aad:2015vsa} while CMS favors a smaller 
one $\mu^{\tau\tau}_{\text{CMS}}=0.78\pm0.27$~\cite{Chatrchyan:2014nva}. We take a $\chi^2$ analysis at 95\% C.L. for these two 
measurements, assuming a Gaussian distribution for both and neglecting 
their correlations. Apparently, the allowed parameter region should be a circular band at the ($\text{Re}(y_{\tau})$, $\text{Im}(y_{\tau})$) plane, 
as is indicated by two green dashed curves in Fig.~\ref{CPVTauTau}.
A future determination of this coupling that agrees with the SM value within $\pm 10\%$ 
is plotted as a curved blue band. 

\noindent{\bfseries $\bm{h\rightarrow\tau\mu}$ constraints.}
The lepton flavor-changing decay width is given by
\begin{eqnarray}
  \Gamma^{\tau\mu} = \frac{\sqrt{2}c_{\beta-\alpha}^2 G_F m_h}{8\pi } 
                     |N_{\tau\mu}|^2 \ \ .
  \label{}
\end{eqnarray}
Theoretically, a sizable $\text{Br}(h\rightarrow \tau\mu)$ requires a small $|M_{32}|$ (see Eq. (\ref{NEtm})). 
ATLAS sets an upper limit on its branching ratio, $\text{Br}(h\rightarrow \tau\mu) < 1.43\%$,  
at 95\% C.L.~\cite{Aad:2016blu}, while CMS gives a best fit 
$\text{Br}(h\rightarrow \tau\mu) = 0.84^{+0.39}_{-0.37}\%$ as well as an upper limit 
$\text{Br}(h\rightarrow \tau\mu) < 1.51\%$ at 95\% C.L.~\cite{Khachatryan:2015kon}.  
In Fig.~\ref{CPVTauTau}, the current ATLAS upper limit $1.43\%$ as well as two projected ones, say,  $1\%$ and $0.5\%$, are shown as dashed curves, in the two cases with $r_{32}=0.9$ and $1.1$. The circular boundaries of the brown regions correspond to vanishing $M_{33}$ or $N_{\tau\mu}$, yielding $\text{Br}(h\rightarrow \tau\mu) =0$.

\noindent{\bfseries $\bm{\tau\rightarrow\mu\gamma}$ constraints.}
Non-vanishing $N_{\tau\mu}$ may also contribute to the rare decay $\tau\rightarrow \mu\gamma$, via one-loop neutral and charged Higgs mediated diagrams and 
two-loop Barr-Zee type diagrams~\cite{Barr:1990vd,Chang:1993kw}. Explicitly, one has  
\begin{eqnarray}
  \text{Br} (\tau \rightarrow \mu\gamma) =  \frac{\tau_\tau \alpha G_F^2 m_{\tau}^5}{32 \pi^4} ( |C_{7L}|^2 + |C_{7R}^2 | ),
  \label{}
\end{eqnarray}
where $\tau_{\tau} = (290.3\pm 0.5) \times 10^{-15}s$~\cite{Agashe:2014kda} is the $\tau$ lifetime  and 
$C_{7L/R}$ are the Wilson coefficients of the dipole operators
\begin{eqnarray}
Q_{7}^{L/R} = \frac{e}{8\pi^2} m_{\tau} \bar{\mu} \sigma^{\mu\nu} (1\mp \gamma^5) \tau F_{\mu\nu},
  \label{DipoleOperators}
\end{eqnarray}
in the Hamiltonian
$-G_F [ C_{7L} Q_7^{L} +  C_{7R} Q_7^{R} ]/\sqrt{2}$\cite{Buras:1998raa}. In our setup, $C_{7L}$ and $C_{7R}$ are proportional to $N^*_{\tau\mu}$ and $N_{\mu\tau}$, respectively,  yielding a vanishing $C_{7R}$. The current experimental limit is $\text{Br}(\tau\rightarrow \mu\gamma) < 4.4\times 10^{-8}$ (90\% C.L.)~\cite{Aubert:2009ag}. The allowed parameter regions are denoted in gray in Fig.~\ref{CPVTauTau}. Obviously there exists a positive correlation between the experimental constraints from  $\text{Br}(h\rightarrow \tau\mu)$ and $\text{Br}(\tau\rightarrow \mu\gamma)$, though the relevant new physics contributions result from tree- and loop-levels, respectively.

\noindent{\bfseries  Electric dipole moments.} Null results from experimental searches for the electric dipole moments (EDMs) of the neutron, neutral atoms, and molecules  in general place stringent limits on new sources of CPV. In the present instance, the electron EDM ($d_e$) provides the most significant probe of $\Im (J_A)$ or $\Im y_\tau$, given the bound obtained by the ACME collaboration using ThO\cite{Baron:2013eja}. In our setup, the dominant contribution to electron EDM results from $h-$mediated Barr-Zee diagram with a $\tau$ lepton loop, because of non-vanishing $\Im y_\tau$.  We find $|d_e/e|\approx 1.66\times 10^{-29}\, |\Im y_\tau|\text{cm}$, yielding a bound of $|\mathrm{Im} y_\tau| < 5.2$. As indicated in Fig.~\ref{CPVTauTau}, this bound is an order of magnitude larger than what is required to account for the observed BAU (see below). We also note in passing that CPV in the scalar potential, associated with a different Jarlskog invariant, will lead to mixing between the CP-even and CP-odd scalars. The resulting EDM contributions can be considerably larger (for a given value of the relevant Jarlskog invariant). For a recent analysis, see Ref.~\cite{Inoue:2014nva}.

\noindent{\bfseries Electroweak baryogenesis.} The first order EWPT proceeds via bubble nucleation. CPV scattering from the bubble walls generates a net left-handed fermion density $n_L$, whose diffusion ahead of the advancing wall biases the electroweak sphalerons into producing a net baryon number density, $n_B$. The expanding bubbles capture and preserve this density if the sphaleron processes are sufficiently quenched inside the bubbles. We compute $n_L$ from a set of quantum transport equations, derived from the equations of motion for Wightman functions arising in the closed time path formulation of non-equilibrium quantum field theory by expanding in gradients of the bubble wall profile and chemical potentials (see Ref.~\cite{Lee:2004we} for pedagogical discussions). As with earlier work, we will employ the \lq\lq vev insertion approximation", which provides a reasonable estimation of the CPV sources (see Ref.~\cite{Morrissey:2012db} for a discussion of theoretical issues associated with the computation of these sources), and work in fermion weak basis (i.e. approximate fermion mass basis during EWPT) in this section. Since the weak sphaleron rate $\Gamma_\mathrm{ws}$~\cite{Bodeker:1999gx} is much smaller than the rates for diffusion and particle number changing reactions that govern $n_L$~\cite{Chung:2008aya}, we first solve for this density and substitute the result into the equation for $n_B$. 

For simplicity, we neglect bubble wall curvature~\cite{Cline:2000nw}, so that  
the quantities entering the quantum transport equations depend only on 
the coordinate in the bubble wall rest frame $\bar{z} = z + v_w t$ with  $v_w$ being the wall velocity, $\bar{z}>0$
corresponding to broken phase and $\bar{z}<0$ for unbroken phase.
Since non-zero densities for the first 
and second generation quarks as well as for the bottom quark are generated only by strong sphaleron 
processes, the following relations hold:  $Q_1 = Q_2 = -2 U = -2 D =-2 C = -2S = - 2B$, where $Q_k$ denotes the density of left-handed quarks of generation $k$ and $U$, $D$, {\em etc.} denote the corresponding right-handed quark densities.
In addition,  $L_1=L_2=e_R\approx 0$ since the corresponding leptonic Yukawa interactions are negligible compared to those retained in our choice of Yukawa texture. Local baryon number density is also 
approximately conserved on the time scales relevant to the reactions that govern $n_L$, so that $\sum_{i=1}^3 (Q_i + U_i + D_i) =0$. The resulting transport equations are
\begin{eqnarray}
&&\partial_{\mu} Q_3^{\mu}     = \Gamma_{mt} (\xi_T-\xi_{Q_3}) 
                                + \Gamma_t (\xi_{T}-\xi_H-\xi_{Q_3}) \nonumber \\
&&    \hspace{1.3cm}            + 2 \Gamma_{ss} \delta_{ss}, \nonumber\\
&&\partial_{\mu} H             = \Gamma_t (\xi_{T}-\xi_H-\xi_{Q_3}) 
			        +\Gamma_{\tau}(\xi_{L_3}-\xi_{\tau_R}-\xi_H) \nonumber \\
&&    \hspace{1.2cm}           -2 \Gamma_h \xi_H, \nonumber\\
&&\partial_{\mu} L_3^{\mu}     =- \Gamma_{m \tau}(\xi_{L_3}-\xi_{\tau_R}) 
			        -\Gamma_{\tau}(\xi_{L_3}-\xi_{\tau_R}-\xi_H) \nonumber \\
&& 		    \hspace{1.4cm}		        + S^{CPV}_{\tau_L},  \nonumber \\
&&\partial_{\mu}{\tau_R}^{\mu} =-\Gamma_{\tau} (\xi_H + \xi_{\tau_R}-\xi_{L_3}) 
			        +\Gamma_{m\tau}(\xi_{L_3}-\xi_{\tau_R}), \nonumber \\
&&\partial_{\mu} T^{\mu}       =-\Gamma_{mt} (\xi_T-\xi_{Q_3}) 
				-\Gamma_t (\xi_{T}-\xi_H-\xi_{Q_3}) \nonumber \\
&& 		    \hspace{1.4cm}			-\Gamma_{ss} \delta_{ss}, \nonumber \\
&&\partial_{\mu} \mu_R^{\mu}   =S_{\mu_R}^{CPV},
  \label{eq:transport}
\end{eqnarray} 
where $\delta_{ss} =\xi_T + 9 \xi_B -2 \xi_{Q_3}$, $\xi_a = n_a/k_a$, with $k_a$ being the statistical 
weight~\cite{Lee:2004we} associated with the number density $n_a$ of species ``a'' and 
$\partial_{\mu} \approx v_w \frac{d}{d \bar{z}}-D_a \frac{\partial^2}{d \bar{z}^2}$  
with $D_a$ being the diffusion constant~\cite{Joyce:1994zn} from the diffusion approximation.
The CPV source terms are 
\begin{eqnarray}
  S^{CPV}_{\tau_L} &=& - S^{CPV}_{\mu_R} = 
  \frac{ v^2(\bar{z}) v_w \frac{d{\beta(\bar{z})}}{d{\bar{z}}} { \text{Im}(J_A)}}{2\pi^2} \, \mathcal{I}\ \ ,
  \label{eq:source}
\end{eqnarray}
where $\mathcal{I}$ is a momentum-space integral that depends on the leptonic thermal masses (see Ref.~\cite{Liu:2011jh}) and $d\beta/d{\bar z}$ characterizes the local variation of $\tan\beta({\bar z})$ as one moves across the bubble wall. 
Furthermore
$\Gamma_{ss} \approx 16 \alpha_s^4 T$ is the strong sphaleron rate~\cite{Giudice:1993bb}; 
$\Gamma_{mt}$ is the two body top relaxation rate~\cite{Lee:2004we}; and 
$\Gamma_{t/\tau}$ is the $t/\tau$ Yukawa induced three body rate~\cite{Cirigliano:2006wh}.
After solving for the densities in Eqs.~(\ref{eq:transport}), we obtain 
$n_L = \sum_{i}(Q_i + L_i)$~\cite{Carena:2002ss} and $n_B$, which is a constant in the 
broken phase:
\begin{eqnarray}
  n_B = \frac{3 \Gamma_\mathrm{ws}}{D_q \lambda_+} \int_0^{-\infty}n_L(\bar{z}) e^{-\lambda_- \bar{z}} d\bar{z}\ \  ,
\end{eqnarray}
where $\Gamma_{\text{ws}} \approx 120 \alpha_w^5 T$~\cite{Bodeker:1999gx} 
and $\lambda_{\pm} = (v_w \pm \sqrt{v_w^2 + 15 \Gamma_{\text{ws}}D_q})/(2 D_q)$.

Assuming a fast $\tau_R$ diffusion~\cite{Chung:2009cb}, we solve the transport equations perturbatively at the leading order 
of $\Gamma_{t}^{-1}$, $\Gamma_y^{-1}$, $\Gamma_{\tau}^{-1}$ and $\Gamma_{ss}^{-1}$. We have further neglected
$\Gamma_{m\tau}$ in the final result as it is generally small compared with $\Gamma_{mt}$; then $n_B$ is proportional to 
$\text{Im} (y_{\tau})$ with no dependence on $\text{Re}(y_{\tau})$. One important remaining parametric uncertainty is the 
difference of $\beta(\bar{z})$ in the broken and symmetric phases ($\equiv \Delta \beta$) since 
the CPV source term and thus $n_B$ are both directly proportional to it. Here we take its maximum magnitude to be $0.4$ and vary it to obtain the bands in Fig.~\ref{CPVTauTau}
where the upper and lower bands give opposite signs of BAU resulting from the unknown sign of $\Delta \beta$. Imposing the condition $|M_{32}| < m_\tau$ as discussed above then restricts $\text{Re}(y_{\tau})$ to the region of overlap between the pink bands and the two circular regions.

\noindent{\bfseries Results and collider probes.} 
Combining the analyses above, we  find that there exist  parameter regions in Fig.~\ref{CPVTauTau} where the observed BAU can be explained without violating current experimental bounds. These regions are characterized by $|\Im(y_\tau)| \gtrsim \mathcal{O}(0.1)$, corresponding to $|\Im (J_A)|\gtrsim \mathcal{O}(10^{-5})$, or $|\phi_\tau| > \mathcal{O}(10{\degree})$. 
As indicated above, the present EDM upper bounds on these CPV parameters are roughly an order of magnitude larger than the BAU requirements. The next generation searches for neutron, atomic, and molecular EDMs that plan for order of magnitude or better improvements in sensitivities may, thus, begin to probe the BAU-viable parameter space.

Alternatively, collider measurements of the CP properties of the $h\bar{\tau}\tau$  coupling may also test this scenario. 
For example, a recent study shows that use of the $\rho$-meson decay plane method or impact parameter method at the LHC may allow a determination of  $\phi_{\tau}$ with an uncertainty of $15{\degree}(9{\degree})$  with an integrated luminosity of $150 \text{fb}^{-1}$($500\text{fb}^{-1}$), or $\sim 4{\degree}$ with $3\,  \text{ab}^{-1}$~\cite{Berge:2015nua}. At Higgs factories, $\phi_{\tau}$ could be measured with an accuracy $\sim 4.4{\degree}$, with a $250\, \text{GeV}$ run and $1\, \text{ab}^{-1}$ luminosity~\cite{Harnik:2013aja}. Therefore, the collider measurements of the CP-properties of the $h\bar{\tau}\tau$  coupling complement the measurements of $h\to \tau\mu$ or $\tau\to \mu \gamma$, which constrain more the parameter regions with relatively small $|\Im y_\tau|$, or $|\phi_\tau|$.

\noindent{\bfseries Discussion.} 
We stress that $N_{\tau\mu}$ or $N_{\mu\tau}$ are not involved in any Jarlskog invariants here. Indeed, for the Yukawa texture considered in this study, we have $N_{\mu\mu} = N_{\mu\tau}=0$.  Consequently, there is only one independent phase in the $\tau-\mu$ system (apart from an overall common phase), with the corresponding Jarlskog invariant defined by the imaginary part of Eq. (\ref{Jinv}). This implies that CP-violation in this setup does not affect $y_{\tau\mu}$ or $y_{\mu\tau}$.  

With a more generalized texture, the situation can be different~\cite{Botella:1994cs}. For example, one can define a new Jarlskog invariant~\cite{Botella:1994cs}
\begin{eqnarray}
\Im (J_B) \sim \Im (\lambda_7^{\rm HB}  N_{\mu\mu}  N_{\tau\mu}  N_{\mu\tau}  m_{\tau})
\end{eqnarray}
with $\lambda_7^{\rm HB}$ being the coefficient of the term $H_2^\dagger H_2 H_1^\dagger H_2$ in Higgs potential.   
Here the value of $N_{\mu\mu}$ need not be small, since $H_2$ does not directly contribute to SM fermion mass generation. Nonetheless, 
the effects of $\Im (J_B)$ on both $T=0$ phenomenology and CPV dynamics at finite $T$ arise a loop level, 
since four instead of two Yukawa couplings (or fermion mass parameters) are involved. 
Explicitly, $\Im (J_B)$ can be probed  
via the interference between the tree-level and one-loop diagrams in the decay of $h \to \mu \tau$. Similarly, $\Im (J_B)$ can contribute 
the BAU generation at one-loop level, as it enters the one-loop, self-energy correction of fermions that contribute to spacetime-dependent vev scattering,  yielding a loop-level CPV source. For similar reasons, its contribution to muon EDM typically requires 
extra mass insertions, hence being negligibly small. A full exploration in this regard will be deferred to future work.

\noindent{\bfseries [Note added]} While this article was being finalized, the paper~\cite{Chiang:2016vgf} appeared, which partially overlaps with this one in discussing the correlation between electroweak baryogenesis and the $\tau-\mu$ Yukawa structure in the 2HDM. However, in contrast to Ref.~\cite{Chiang:2016vgf},  we having explicitly show how the baryon asymmetry generated intrinsically correlates with the CP-violating decay $h\to \tau\tau$, using a Jarlskog invariant uniquely defined at tree level.  As for the complex phases of Higgs couplings with $\tau$ and $\mu$ leptons, which were addressed in Ref.~\cite{Chiang:2016vgf}, we note that they contribute only to Jarlskog invariants defined as a product of at least four Yukawa couplings (or fermion mass parameters), and hence play a sub-leading role in this exploration generically.  

\noindent{\bfseries Acknowledgements.} 
We would like to thank W. Chao and C. Y. Seng for helpful discussions. MJRM and HKG are supported in part under U.S. Department of Energy contract DE-SC0011095. 
HKG is also supported by the China Scholarship Council. JS is supported by the Strategic Priority Research Program of the Chinese Academy of Sciences
under Grant No. XDB21010200. YYL is supported by the the Hong Kong PhD Fellowship Scheme (HKPFS).  TL is supported by the Collaborative Research Fund (CRF) under Grant No.  HUKST4/CRF/13G and the General Research Fund (GRF) under Grant No. 16312716. Both the HKPFS and the CRF, GRF grants are issued by the Research Grants Council of Hong Kong S.A.R..  MJRM, TL and JS would like to thank the hospitality of MIAPP (TL would extend the thanks to MITP and Aspen Center for Physics) during the finalization of this article. 
\appendix


\begin{thebibliography}{99}
\bibitem{CMS:13Tev}
  Preliminary 13 TeV CMS results provide no indication of a non-zero CLFV Higgs decay, although the significance is not sufficient to preclude a non-vanishing
  $h\tau\mu$ coupling.
\end{thebibliography}
\end{document}